# The Role of Low-lying Optical Phonons in Lattice Thermal Conductance of Rare-earth Pyrochlores: A First-principle Study

Guoqiang Lan[a], Bin Ouyang[a], Jun Song[a, 1]

[a] *Department of Mining and Materials Engineering, McGill, University, Montreal, Quebec H3A 0C5, Canada*

**Abstract**

Rare-earth pyrochlores, commonly exhibiting anomalously low lattice thermal conductivities, are considered as promising topcoat materials for thermal barrier coatings. However the structural origin underlying their low thermal conductivities remain unclear. In the present study, we investigated the phonon properties of two groups of RE pyrochlores, $Ln_2Zr_2O_7$ (Ln = La, Nd, Sm, Gd) and $Gd_2T_2O_7$ (T = Zr, Hf, Sn, Pb) employing density functional theory and quasi harmonic approximation. Through the relaxation time approximation (RTA) with Debye model, the thermal conductivities of those RE pyrochlores were predicted, showing good agreement with experimental measurements. The low thermal conductivities of RE pyrochlores were shown to largely come from the interference between the low-lying optical branches and acoustic branches. The structural origin underlying the low-lying optical branches was then clarified and the competition between scattering processes in transverse and longitude acoustic branches was discussed.

**Keywords:** rare-earth, pyrochlores, phonon dispersion, thermal conductivity,

[1]Author to whom correspondence should be addressed. Email: jun.song2@mcgill.ca

density functional theory

# 1. Introduction

Thermal barrier coating (TBC) [1-3] is an essential constituent in gas turbine engines. It is applied to surfaces of superalloy components in the engine, shielding them from the hot gas to allow an operating temperature much higher than their melting temperatures. The thermal insulation from TBC is determined by the outmost ceramic topcoat [1-3]. Currently the common material used for the top coat is 7wt.% yttria stabilized zirconia (7YSZ) because of its relatively low thermal conductivity (2.0~3.0 $Wm^{-1}K^{-1}$) [4, 5], good toughness [6, 7] and thermochemical compatibility with the underlying TGO [4, 8]. However, 7YSZ suffers from accelerated sintering kinetics [9, 10] and destabilization of the desirable the metastable $t'$ phase at elevated temperatures [11], and susceptibility to calcium-magnesium aluminosilicate (CMAS) penetration [12, 13], which significantly degrades its insulating efficiency, strain tolerance and intrinsic toughness, being greatly detrimental to its performance and durability. These limitations necessitate innovation to develop alternative coating materials.

Rare-earth (RE) pyrochlores recently emerge as potential candidate materials for the next-generation top coat given their low intrinsic thermal conductivities (i.e., 1~2 $Wm^{-1}K^{-1}$) [14-19]. The RE pyrochlore also exhibits high temperature stability [4], similar thermal expansion coefficient as its beneath substrate [20, 21], and can react with the CMAS melt to seal the surface against further infiltration [22]. In addition, RE pyrochlores can well preserve the pore content and architecture owing to their sluggish sintering kinetics [23]. The thermal properties of various RE pyrochlores were well characterized in many experimental studies [14-19]. It was postulated in some studies [24, 25] that the abnormally low thermal conductivities of these pyrochlores may be attributed to the abundance of intrinsic oxygen vacancies and low-lying optical phonons due to RE elements rattling. However the exact structural origin and phonon characteristics, particularly at the atomic level, remains largely unknown. Recently a few studies attempted to model and understand the thermal transport in pyrochlores. Liu et al. [26] and Feng et al. [27, 28] investigated the thermal conductivities of $Ln_2Zr_2O_7$, $Ln_2Sn_2O_7$ (Ln = La, Nd, Sm, Gd) and $La_2T_2O_7$ (T = Ge, Ti, Sn Zr, Hf) systems using density functional theory (DFT). In their studies, several physical properties, including elastic constants, lattice spacing and density, were computed from first-principles calculations, and subsequently combined with the simple formulas suggested by Clarke [29], Slack [30] and Cahill et al. [31] to estimate the minimum thermal conductivity. The estimated values yield good ball-park agreements with experimental measurements. Nonetheless, those studies fail to distinguish different RE pyrochlores in term of thermal conductivity and thus do not reveal the mechanistic origin of lattice heat conduction.

In this paper, we employ the approach of relaxation-time approximation (RTA) together with the Debye model [32] to study the high-temperature thermal conductivities of several RE pyrochlores. Different from the simple models invoked in those previous modeling studies of pyrochlore oxides [26-28, 33], this approach allows direct considerations of phonon scattering processes and anharmonic relaxation times, and is evidenced by a few studies to an accurate method to predict the thermal conductivity for a wide variety of solid-state materials [32, 34-36]. We focused on two sets of RE pyrochlores, $Ln_2Zr_2O_7$ (Ln = La, Nd, Sm, Gd) and $Gd_2T_2O_7$ (T = Zr, Hf, Sn, Pb). Through first-principle DFT calculations, the ground-state crystal structures and corresponding elastic constants of those pyrochlores were obtained. Meanwhile the harmonic and anharmonic vibrational properties, including phonon density of states (DOS), phonon dispersion and Grüneisen parameters, are thoroughly investigated by DFT phonon calculations within the quasi harmonic approximation (QHA) [37]. Using the information obtained from DFT, the thermal conductivities of the RE pyrochlores considered were evaluated within RTA along with the Debye approximation, and compared with the ones estimated from Clarke's and Cahill's models. The predictions from RTA along with the Debye approximation were demonstrated to effectively differential between different RE pyrochlores and yield much more consistent agreement with experimental data. Finally, the atomic vibration patterns contributing to the low-lying optical modes that scatter the acoustic modes were investigated to clarify the mechanistic origin of the low thermal conductivity of RE pyrochlores.

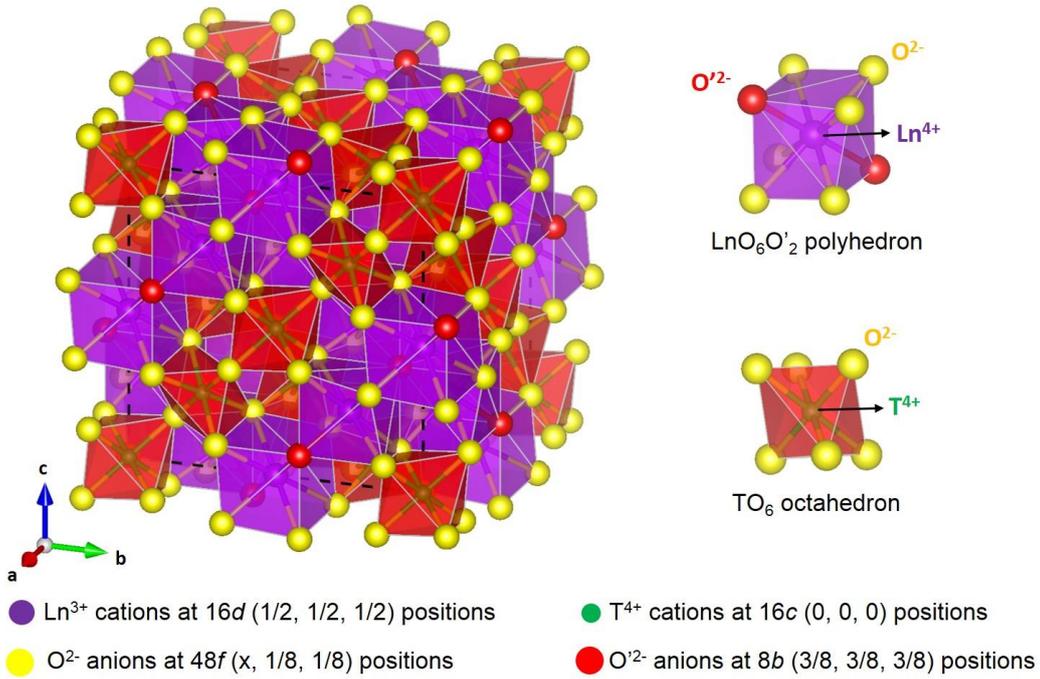

**Fig. 1** (Color online): Crystal structure of $Ln_2T_2O_7$ type pyrochlore (with the space group being $F m\bar{3} m$ and a chemical formula as $Ln_2T_2O_6O'$). The structure are occupied by four crystallographically unique atomic sites [38]: larger $Ln^{4+}$ cations at 16$d$, smaller $T^{4+}$ cations at 16$c$, $O^{2-}$ anions at 48$f$ and $O'^{2-}$ anions at 8$b$. Parameter $x$ is used to represent the off-center property of O anion at 48$f$ position. The two polyhedral, i.e., $LnO_6O'_2$ and $TO_6$, that compose the pyrochlore are illustrated by the two subfigures on the right.

## 2. Methodology

### 2.1. Density functional theory and phonon calculations

DFT calculations were performed using the Vienna ab initio simulation package (VASP) [39, 40] with Perdew–Burke–Ernzerhof [41] GGA+U approach [42, 43] based on plane-wave basis sets. The values of the Hubbard energy U were set following Ref. [27]. Valence configurations of the elements used in this study are the follows: La-$5s^2\ 5p^6\ 5d^1\ 6s^2$, Nd-$4f^4\ 5s^2\ 5p^6\ 6s^2$, Sm-$4f^6\ 5s^2\ 5p^6\ 6s^2$, Gd-$4f^7\ 5s^2\ 5p^6\ 5d^1\ 6s^2$, Zr-$4s^2\ 4p^6\ 5d^2\ 5s^2$, Hf-$5s^2\ 5p^6\ 5d^2\ 6s^2$, Sn-$5s^2\ 5p^2$, Pb-$6s^2\ 6p^2$, and O-$2s^2\ 2p^4$. The electron–core interaction was described by the Blöchl's projector augmented wave method (PAW) within the frozen-core approximation [44]. For the structural optimization and computation of elastic constants, a conventional pyrochlore cubic unit cell containing 88 atoms was constructed, and a 2×2×2 Monkhorst-Pack (MP) k–mesh for Brillouin-zone integrations and a cutoff energy of 450 eV for the plane-wave basis set were used in the DFT calculations with convergence criteria for energy and force on each ion set as $10^{-9}$ eV and $10^{-6}$ eV Å$^{-1}$, respectively. The real-space force constants were obtained with the density-functional perturbation theory (DFPT) [45] using a 2×2×2 supercells (176 atoms) constructed from the pyrochlore primitive cell (22 atoms). The phonon properties were evaluated using PHONOPY [46] on 23×23×23 $q$ grid for all the RE pyrochlore systems. In the phonon calculations within QHA, the cell volume was varied by ±0.75% with respect to the equilibrium volume.

**2.2. Models for lattice thermal conductivity**

*2.2.1. Minimum thermal conductivity based on Clarke's and Cahill's model*

The concept of minimum thermal conductivity, $\kappa_{\min}$, was developed by Slack [47], stating that for a material there is a lower limit for the intrinsic thermal conductivity. It is of great technological importance to the TBC material research and has been widely used to predict the plateau thermal conductivities of poly-ion ceramic oxides at high temperature. Clarke *et al.* [29] proposed an expression of $\kappa_{\min}$, assuming that the phonon mean free path is identical to the inter-atomic spacing and the specific heat approaches $3k_B$ ($k_B$ being the Boltzmann's constant) per atom above the Debye temperature:

$$\kappa_{\min} = 0.87 k_B \bar{M}^{-2/3} E^{1/2} \rho^{1/6}, \tag{1}$$

where $\bar{M}$ is the average mass of per atom, $E$ the Young's modulus and $\rho$ is the density. In the work by Cahill et al. [31], in accordance with Einstein's notion of lattice vibration, they assume that individual oscillators vibrate independently of one another and the phonon scattering time is one half the period of vibration. Under those assumptions, $\kappa_{\min}$ at high temperature limit can be expressed as

$$\kappa_{\min} = \frac{k_B}{2.48} p^{2/3} (v_l + 2v_t), \tag{2}$$

where $p$ is the density of number of atoms per volume, and $v_l$ and $v_t$ are the average longitudinal and transverse sound velocities respectively. As apparently illustrated in Eqs. 1-2, both Clarke's and Cahill's models predict that lower $\kappa_{\min}$ requires smaller modulus (or sound velocity) and larger size of the unit cell.

*2.2.2. Relaxation-time approximation (RTA) with the Debye model*

RTA is commonly invoked in the treatment of Boltzmann transport equation (BTE) [32]. Given that it is usually difficult to consider various phonon scattering relaxation times for the full phonon dispersions, RTA is often augmented by the Debye model to simplify the phonon branches.

Phonon dispersion curves normally consist of low-frequency acoustic and high-frequency optical branches [32]. Generally, only acoustic phonons are effective for heat conduction, whereas optical phonons contribute very little to heat transport [32]. Consequently the total thermal conductivity, $\kappa$ can be simplified as just a sum of contributions from two transverse ($\kappa_{TA}$ and $\kappa_{TA'}$) and one longitudinal ($\kappa_{LA}$) acoustic phonon branches as discussed in Refs [34-36]:

$$\kappa = \kappa_{TA} + \kappa_{TA'} + \kappa_{LA}. \tag{3}$$

These partial conductivities $\kappa_i$ ($i$ = TA, TA' or LA stands for two transverse and one longitudinal acoustic phonon branches, respectively) are given by RTA with Debye approximation [32] as:

$$k_i = \frac{k_B}{2\pi^2 v_i}\left(\frac{k_B T}{\hbar}\right)^3 \int_0^{\tilde{\Theta}_i/T} t_q \frac{x^4 e^x}{(e^x - 1)^2} dx, \tag{4}$$

where $\hbar$ is the reduced Plank constant, $T$ is temperature, $v_i$ is corresponding acoustic phonon velocity, and $x = \hbar\omega/k_B T$ is a dimensionless quantity with $\omega$ being the phonon frequency. $\tilde{\Theta}_i$ is the reduced Debye temperature determined from the conventional Debye temperature, $\Theta_i$, given in Eq. 5 below for a crystal with $n$ atoms per unit cell and the atomic volume being $a^3$ [48-50]

$$\tilde{\Theta}_i = \Theta_i n^{-1/3} \tag{5}$$

$$\Theta_i = \frac{\hbar}{k_B}\left(\frac{6\pi^2}{a^3}\right)^{1/3} v_i \tag{6}$$

The parameter $\tau_q$ represents the total scattering time due to various resistive processes, Umklapp scattering (U), phonon-point-defect scattering (PD), phonon-boundary scattering (B), and *etc.* [32] In perfect insulators at high temperatures, U-processes dominate the phonon scattering process, and thus we have $\tau_q^{-1} \approx \tau_U^{-1}$. Based on the Leibfried and Schlomann model [51], Slack *et al.* [52] suggested the following form for Umklapp scattering rate:

$$\frac{1}{t_U^i(x)} \approx \frac{\hbar \bar{g}_i^2}{\bar{M} v_i^2 \tilde{\Theta}_i} \left(\frac{k_B}{\hbar}\right)^2 x^2 T^3 \exp(-\tilde{\Theta}_i / 3T) \qquad (7)$$

where $\bar{g}_i$ is the average Grüneisen parameter for the $i$ ($i$ = TA, TA' or LA) acoustic branch and $\bar{M}$ is the average mass per atom in the material. Other formulas of the Umklapp scattering rate also exist, all showing $\tau_U^{-1}$ proportional to $\gamma^2$ [48, 53, 54]. The mode Grüneisen parameter can be obtained as [32, 55]

$$\gamma(q,s) = -\frac{V}{\omega(q,s)} \frac{\partial \omega(q,s)}{\partial V}, \qquad (8)$$

where $\gamma(q,s)$ and $\omega(q,s)$ represent the mode Grüneisen parameter and phonon frequency at wave vector $q$ and branch $s$. $\gamma(q,s)$ can be interpreted as the rate of change of $\omega(q,s)$ with unit cell volume $V$. The average Grüneisen parameter $\bar{\gamma}$ for each acoustic branche can be obtained as

$$\bar{g} = \frac{\sum_{q,s} |g(q,s)| C(q,s)}{\sum_{q,s} C(q,s)}, \qquad (9)$$

where $C(q,s)$ mode specific heat capacity at wave vector $q$ and branch $s$, and the absolute value of mode Grüneisen parameter $|\gamma(q,s)|$ is used to avoid cancellation between different modes. All the parameters in the above equations can be obtained directly from first-principle DFT calculations, allowing the estimation of the thermal conductivity without any empirical data inputs. In the context below, RTA with the Debye model is shorted as RTA model for simplicity.

## 3. Results and discussion

### 3.1. Structural parameters and elastic constants

The crystal structure for the RE pyrochlore oxides is illustrated in Fig. 1, where the relative positions of cations (i.e., $Ln^{3+}$ and $T^{4+}$) and anions (i.e., $O^{2-}$ and $O'^{2-}$) are indicated. The corresponding lattice constants and 48$f$ position parameters (i.e., $x$) are optimized via DFT calculations, listed in Table I where the values from experiments [21, 56-60] are also listed for comparison. We note that for all materials considered the theoretical value deviates from the experimental one by less than 1%.

**Table I**: Calculated and experimental values of the lattice constant, $a$ (Å) and 48$f$ position parameter $x$ for $Ln_2Zr_2O_7$ (Ln = La, Nd, Sm and Gd) and $Gd_2T_2O_7$ (T = Pb, Zr, Hf and Sn). The parameter $r_{Ln}/r_T$ (pm) denotes the corresponding $Ln^{3+}/T^{4+}$ cation radius [61].

|  | $La_2Zr_2O_7$ | | $Nd_2Zr_2O_7$ | | $Sm_2Zr_2O_7$ | | $Gd_2Zr_2O_7$ | | $Gd_2Hf_2O_7$ | | $Gd_2Pb_2O_7$ | | $Gd_2Sn_2O_7$ | |
|---|---|---|---|---|---|---|---|---|---|---|---|---|---|---|
| $r_{Ln}/r_T$ | 103.2/72.0 | | 98.3/72.0 | | 95.8/72.0 | | 93.5/72.0 | | 93.5/71.0 | | 93.5/77.5 | | 93.5/69.0 | |
|  | Cal.[a] | Exp.[b] | Cal.[a] | Exp.[c] | Cal.[a] | Exp.[d] | Cal.[a] | Exp.[e] | Cal.[a] | Exp.[f] | Cal.[a] | Exp.[g] | Cal.[a] | Exp.[h] |
| $a$ | 10.83 | 10.78 | 10.74 | 10.67 | 10.66 | 10.58 | 10.60 | 10.53 | 10.55 | 10.49 | 10.83 | 10.72 | 10.54 | 10.45 |
| $x$ | 0.332 | 0.332 | 0.335 | 0.335 | 0.337 | 0.339 | 0.338 | 0.340 | 0.337 |  | 0.348 |  | 0.338 | 0.335 |

[a] This work (GGA +U)  [b] Ref. [56]
[c] Ref. [57]  [d] Ref. [58]  [e] Ref. [57]
[f] Ref. [21]  [g] Ref. [59]  [h] Ref. [60]

Table II lists the values of anisotropic elastic constants and moduli (Voigt averages [62]) obtained from DFT calculations, and the corresponding experimental data available in literature [63-65]. Despite some absence of experimental data, overall we note that the theoretical values are in good agreement with experimental measurements. In particular for $Ln_2Zr_2O_7$, it is worth noting that our present results exhibit better agreement with experimental data than the ones results calculated using

LSDA + U [27]. The good agreement between our DFT predictions and experimental data (cf. Tables I and II) confirms the GGA+U method is suitable for calculations of the RE pyrochlore systems.

Comparing those parameters in Table II among different pyrochlores, we note that they do not show an apparent relationship with the radii of $Ln^{3+}$ or $T^{4+}$ cations. Particularly concerning the calculated values, we see that for the $Ln_2Zr_2O_7$ group, the moduli of $La_2Zr_2O_7$ are the lowest, while those of other pyrochlores (i.e., Ln = Nd, Sm, Gd) are almost the same. For the $Gd_2T_2O_7$ group, $Gd_2Pb_2O_7$ and $Gd_2Hf_2O_7$ exhibit lowest and highest moduli respectively, while $Gd_2Zr_2O_7$ and $Gd_2Sn_2O_7$ have similar moduli.

**Table II:** Calculated and experimental values of anisotropic elastic constants (in GPa), i.e., $C_{11}$, $C_{12}$, $C_{44}$, bulk modulus $B$ (GPa), shear modulus $G$ (GPa), Yong's modulus E (GPa), the Poisson's ratio $v$, and the $B/G$ ratio. The parameter $r_{Ln}/r_T$ (pm) denotes the corresponding $Ln^{3+}/T^{4+}$ cation radius.

|  | $r_{Ln}/r_T$ [e] | Method | $C_{11}$ | $C_{44}$ | $C_{12}$ | B | E | G | $v$ | B/G |
|---|---|---|---|---|---|---|---|---|---|---|
| $La_2Zr_2O_7$ | 103.2/72.0 | Cal.[a] | 266 | 87 | 110 | 162 | 213 | 83 | 0.281 | 1.95 |
|  |  | Exp.[b] |  |  |  |  | 239 |  |  |  |
| $Nd_2Zr_2O_7$ | 98.3/72.0 | Cal.[a] | 279 | 89 | 107 | 164 | 224 | 88 | 0.273 | 1.87 |
|  |  | Exp.[c] |  |  |  |  | 219 | 86 | 0.271 |  |
| $Sm_2Zr_2O_7$ | 95.8/72.0 | Cal.[a] | 286 | 88 | 107 | 167 | 226 | 89 | 0.274 | 1.88 |
|  |  | Exp.[c] |  |  |  |  | 231 | 90 | 0.277 |  |
| $Gd_2Zr_2O_7$ | 93.5/72.0 | Cal.[a] | 289 | 85 | 103 | 165 | 224 | 88 | 0.273 | 1.87 |
|  |  | Exp.[c] |  |  |  | 156 | 205 | 80 | 0.276 | 1.95 |
|  |  | Exp.[d] |  |  |  | 175 | 238 | 89 | 0.283 | 1.97 |
| $Gd_2Hf_2O_7$ | 93.5/71.0 | Cal.[a] | 301 | 94 | 105 | 170 | 242 | 96 | 0.264 | 1.78 |
| $Gd_2Pb_2O_7$ | 93.5/77.5 | Cal.[a] | 248 | 67 | 93 | 145 | 183 | 71 | 0.289 | 2.03 |
| $Gd_2Sn_2O_7$ | 93.5/69.0 | Cal.[a] | 289 | 87 | 107 | 168 | 226 | 89 | 0.275 | 1.89 |

[a] This work  [b] Ref. [63]  [c] Ref. [64]
[d] Ref. [65]  [e] Ref. [61]

## 3.2. Phonon dispersion and vibrational properties

*3.2.1. Phonon dispersion curves and partial density of states*

The calculated phonon dispersion curves along high-symmetry directions in their Brillouin zones (BZ) and partial density of states (PDOS) for $Ln_2Zr_2O_7$ (Ln = La, Nd, Sm, Gd) and $Gd_2T_2O_7$ (T = Zr, Hf, Sn, Pb) are shown in Fig. 2 and Fig. 3 respectively. In all the plots shown in Figs 2-3, we see there are low-lying optical branches (between 1 and 3 THz) lying in the acoustic range. The low-lying optical branches are identified to be primarily attributed to the vibrations of $Ln^{3+}$ cations.

From Fig. 2, we see that the shapes of the phonon dispersions curves and PDOS of the $Ln_2Zr_2O_7$ pyrochlores are in general rather similar. However we can note that for $La_2Zr_2O_7$, the optical branches attributed to $Ln^{3+}$ cation are higher while the high-frequency branches attributed to $O^{2-}$ (i.e., 8~23 THz) and $O'^{2-}$ are shallower, than the ones for other $Ln_2Zr_2O_7$ pyrochlores. One the other hand, the branches corresponding to $Zr^{4+}$ and low-frequency branches attributed to $O^{2-}$ (i.e., 0~6 THz) are statistically invariant. Given that stronger interatomic bonding results in higher frequencies (for atoms of similar masses) [37], the above observation suggests that the interatomic bonds around $La^{3+}$ be stronger than those around $Nd^{3+}$, $Sm^{3+}$, and $Gd^{3+}$ cations while the bonding environment around $Zr^{4+}$ basically remain the same for all $Ln_2Zr_2O_7$ pyrochlores.

For the phonon dispersions curves and PDOS shown in Fig. 3 for the $Gd_2T_2O_7$ pyrochlores, based on the features of the plots we can divide them into two pairs, i.e., ($Gd_2Pb_2O_7$, $Gd_2Sn_2O_7$) and ($Gd_2Zr_2O_7$, $Gd_2Hf_2O_7$). We can note that the phonon dispersions curves and PDOS show similar features for pyrochlores in the same pair, but are noticeably different for pyrochlores in different pairs. The above observation is understandable as Pb and Sn, and Zr and Hf, respectively are in the same element group possessing similar outermost valence configurations, and thereby have similar bonding environments. The radii of T-type cations are as followed, 77.5 pm for $Pb^{4+}$, 72 pm for $Zr^{4+}$, 71 pm for $Hf^{4+}$ and 69 pm for $Sn^{4+}$ [61]. There is no apparent variation in the frequencies with respect to T-type cation radius. Possible reason for this difference is that most T-type cations exhibit nonmetal features, i.e., there is small electronegativity difference between T and O. Thus some of the T-O bonds show complex covalent characteristics rather than simple ionic features, and consequently the bond strength of T-O bonds cannot be simply determined by the cation radius.

Within the ($Gd_2Pb_2O_7$, $Gd_2Sn_2O_7$) pair, $Gd_2Sn_2O_7$ has a much larger modulus than $Gd_2Pb_2O_7$ (Table II) despite the radius of $Sn^{4+}$ being smaller that of $Pb^{4+}$. This implies that the $Gd_2Sn_2O_7$ has higher average interatomic bond strength than $Gd_2Pb_2O_7$. Consequently, compared to the branches of $Gd_2Pb_2O_7$, all the branches of $Gd_2Sn_2O_7$ show a shift towards higher frequencies, except for those attributed to $Gd^{3+}$, which stay essentially unchanged. On the other hand within the ($Gd_2Zr_2O_7$, $Gd_2Hf_2O_7$) pair, the modulus of $Gd_2Hf_2O_7$ is slightly larger than that of $Gd_2Zr_2O_7$. A close examination of the plots in Figs 3a-b reveals that $Gd_2Hf_2O_7$ exhibits higher-frequency O-branches, slightly lower-frequency T-branches, in comparison to, $Gd_2Zr_2O_7$.

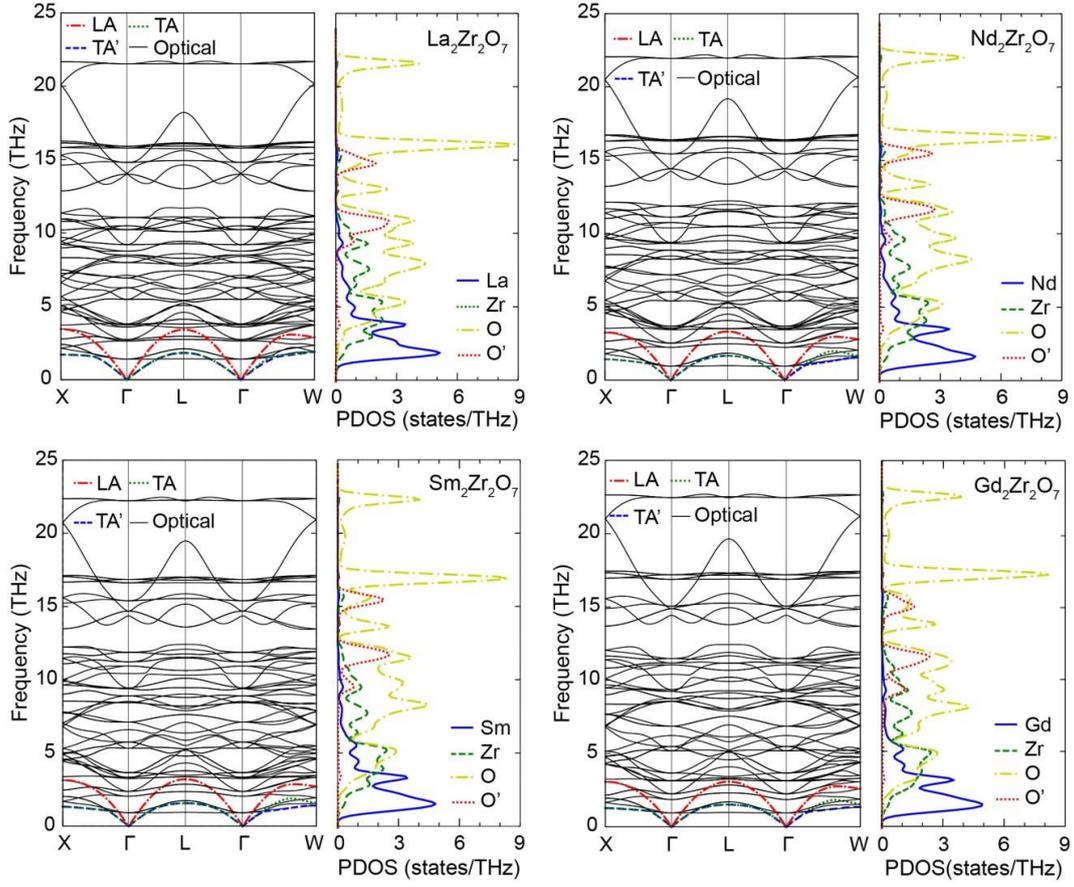

**Fig. 2** (Color online): Phonon dispersions and partial density of states (PDOS) for $Ln_2Zr_2O_7$ (Ln = La, Nd, Sm, Gd) pyrochlores. The two transverse (denoted as TA and TA') and one longitudinal (denoted as LA) acoustic phonon branches are indicated by green dotted, blue dashed and red dashed-dotted lines, respectively, while the optical branches are indicated by solid black lines. For the PDOS, the blue solid, green dashed, yellow dashed-dotted and red dotted curves denote those corresponding to Ln (Ln = Ln, Nd, Sm, Gd), Zr, O and O' atoms, respectively.

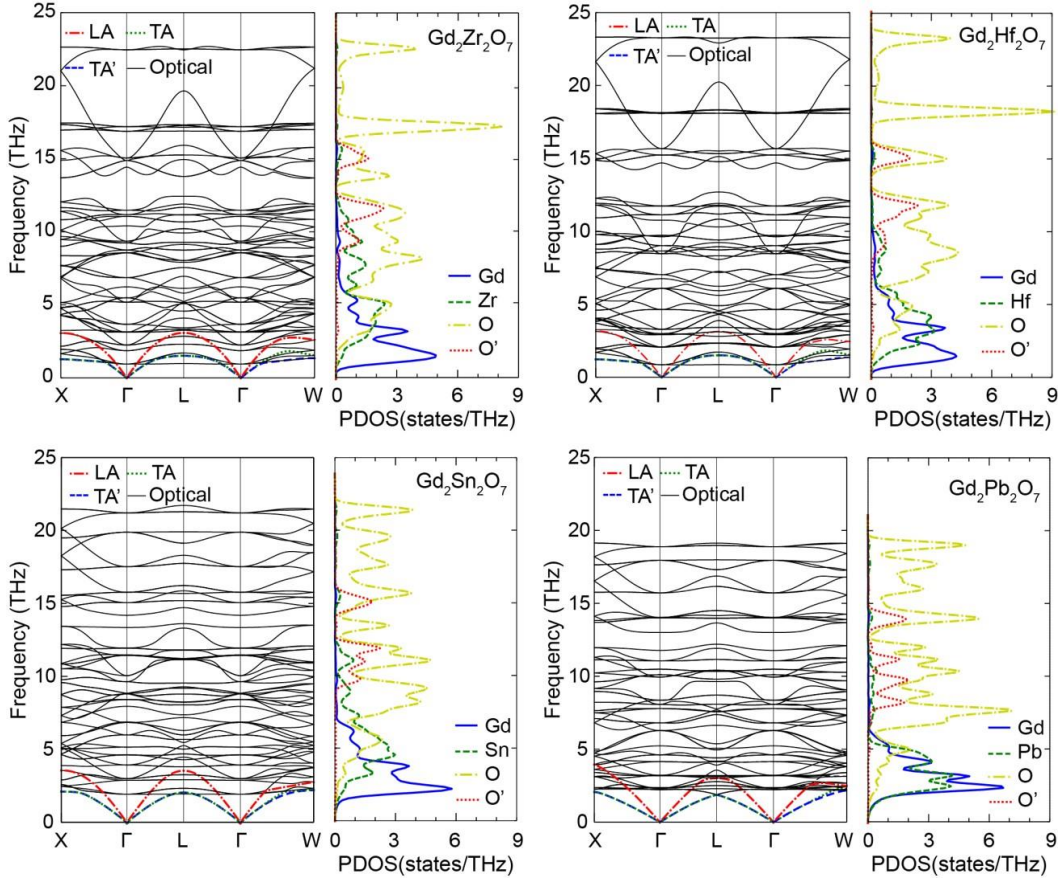

**Fig. 3** (Color online): Phonon dispersions and partial density of states (PDOS) for $Gd_2T_2O_7$ (T = Zr, Hf, Sn, Pb) pyrochlores. The two transverse (denoted as TA and TA') and one longitudinal (denoted as LA) acoustic phonon branches are indicated by green dotted, blue dashed and red dashed-dotted lines, respectively, while the optical branches are indicated by full black lines. For the PDOS, the blue solid, green dashed, yellow dashed-dotted and red dotted curves denote those corresponding to Gd, T (T = Zr, Hf, Sn, Pb), O and O' atoms, respectively.

*3.2.2. Grüneisen parameters*

Using Eq. 8, we computed the dispersion of the Grüneisen parameters ($\gamma$) for all acoustic modes. The phonon dispersions, zoomed in for acoustic branches, and corresponding acoustic Grüneisen dispersion curves are shown in Fig. 4 and Fig. 5 $Ln_2Zr_2O_7$ and $Gd_2T_2O_7$ pyrochlores respectively. There are some general features we can note for both figures. We see the acoustic branches overlap with low-lying optical branches (also previously noted in Section 3.2.1). This results in mixing branches and nonlinear acoustic dispersion curves away from Γ point (the center of the Brillouin zone) [66], and leads to abrupt changes in $\gamma$ due to avoided crossings between optical and acoustic branches (mostly TA branches). Besides the abrupt changes, there are also some small bumps along the Grüneisen dispersion curves (mostly for the LA branches), attributing to cases where acoustic branches cross the optical branches without losing linear features.

Examining the plots for $Ln_2Zr_2O_7$ pyrochlores shown in Fig. 4, we see that the acoustic phonon and Grüneisen dispersion curves are overall similar for different pyrochlores, except for that the TA and TA' branches of $La_2Zr_2O_7$ essentially overlap with each other whereas those of other zirconates diverse at one third of path from Γ to $W$ point. In addition, we can see that the acoustic cut-off frequencies at the zone boundaries decrease as the size of the $Ln^{3+}$ cation decrease. Extremely high $\gamma$ values (above ~10) appear at the TA/TA' branches in $Ln_2Zr_2O_7$ pyrochlores, indicating that TA/TA' branches are strongly scattered by low-lying optical branches and thereby do not contribute much to thermal conduction.

For the set of $Gd_2T_2O_7$ pyrochlores (cf. Fig. 5), we see that $Gd_2Hf_2O_7$ and $Gd_2Zr_2O_7$ exhibit similar dispersion curves. Like $Ln_2Zr_2O_7$ pyrochlores, the Grüneisen dispersion curves of $Gd_2Hf_2O_7$ and $Gd_2Zr_2O_7$ show abrupt changes in $\gamma$ and extremely high values of $\gamma$ at the TA/TA' branches where avoided crossings occur. In comparison, $Gd_2Zr_2O_7$ and $Gd_2Hf_2O_7$ exhibit much less variation in $\gamma$ for the TA/TA' branches. This is because that the lowest optical branches of $Gd_2Pb_2O_7$ and $Gd_2Sn_2O_7$ are located at ~2 THz, higher than those of $Gd_2Zr_2O_7$ and $Gd_2Hf_2O_7$ (located at ~1 THz), so that their TA/TA' branches are able to better retain acoustic feature even near the boundary of Brillouin zone, showing smaller $\gamma$ values. Fundamentally the difference between the ($Gd_2Zr_2O_7$, $Gd_2Hf_2O_7$) and ($Gd_2Pb_2O_7$, $Gd_2Sn_2O_7$) pairs can be attributed to the fact that $Gd_2Zr_2O_7$ and $Gd_2Hf_2O_7$ exhibit strong ionic characters while $Gd_2Pb_2O_7$ and $Gd_2Sn_2O_7$ are of more covalent bonding features. On the other hand, the elevation of the low-lying optical branches in the cases of $Gd_2Pb_2O_7$ and $Gd_2Sn_2O_7$ seemingly leads to more interference between those optical branches with the LA branch. This is particularly noticeable in the case of $Gd_2Sn_2O_7$ where we can note a clear deflection in the LA branch between $\Gamma$ and W that causes an abrupt change in $\gamma_{LA}$ (cf. Fig. 5).

From the phonon and Grüneisen dispersion curves, the average Grüneisen parameters $\bar{g}_i$ ($i$ = TA, TA' or LA) (using Eq. 9), along with the phonon velocities and reduced Debye temperatures, are then calculated for each acoustic branches, shown in Table III. We can see that for the $Ln_2Zr_2O_7$ group and $Gd_2Hf_2O_7$, the TA and TA' branches exhibit much larger values of the average Grüneisen parameter than the LA branch, while for $Gd_2Pb_2O_7$ and $Gd_2Sn_2O_7$ the three acoustic phonon branches show comparable values of the average Grüneisen parameter. Another observation from Table III is that $La_2Zr_2O_7$ exhibits significantly lower $\bar{g}_{TA}/\bar{g}_{TA'}$ than other $Ln_2Zr_2O_7$ pyrochlores.

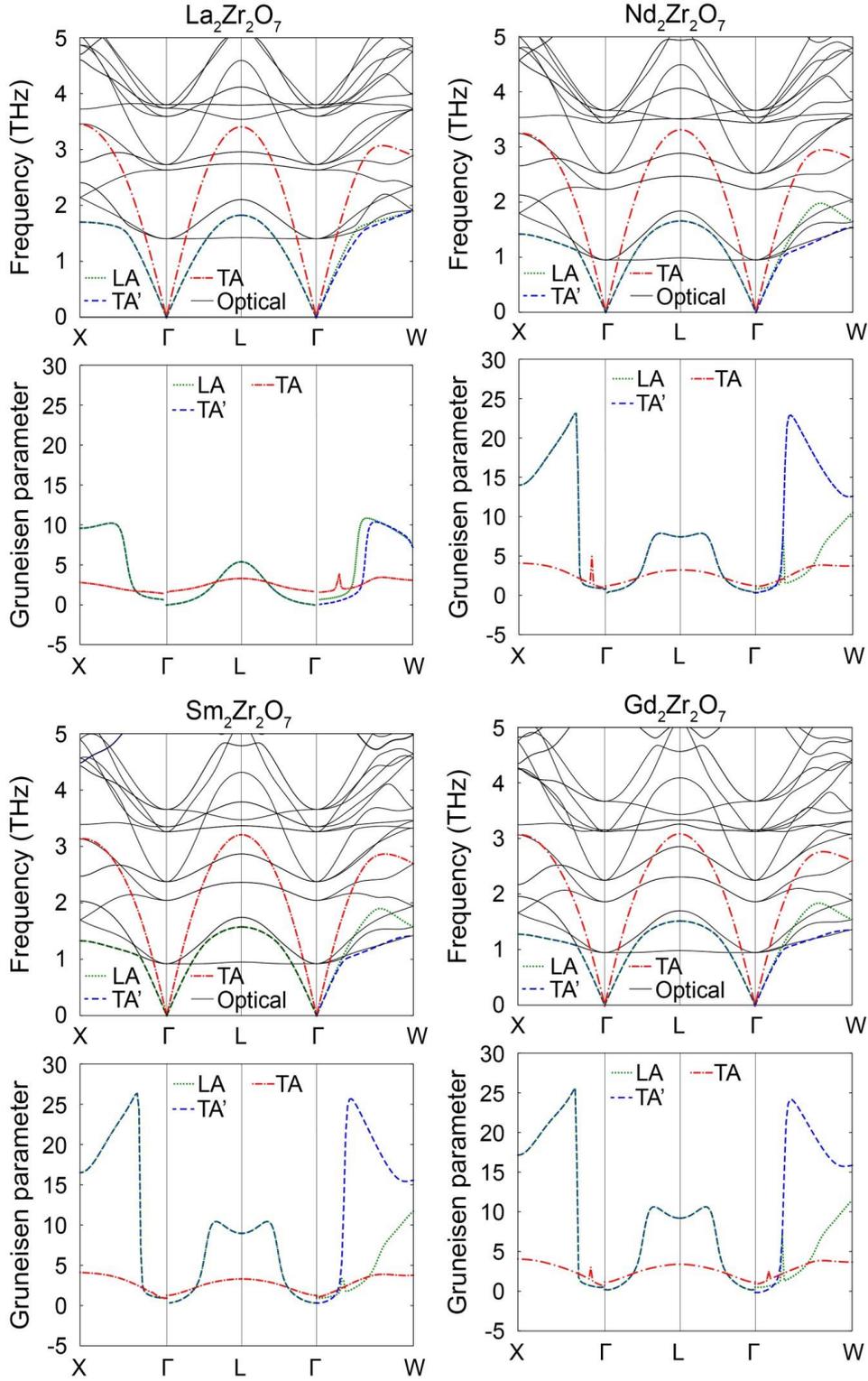

**Fig. 4** (Color online): Calculated phonon dispersions (zoom in for the acoustic branches, i.e., TA, TA',LA and optical branches indicated by green dotted, blue dashed, red dashed-dotted and solid black lines, respectively) and corresponding acoustic Grüneisen parameters for $Ln_2Zr_2O_7$ (Ln = La, Nd, Sm, Gd) pyrochlores.

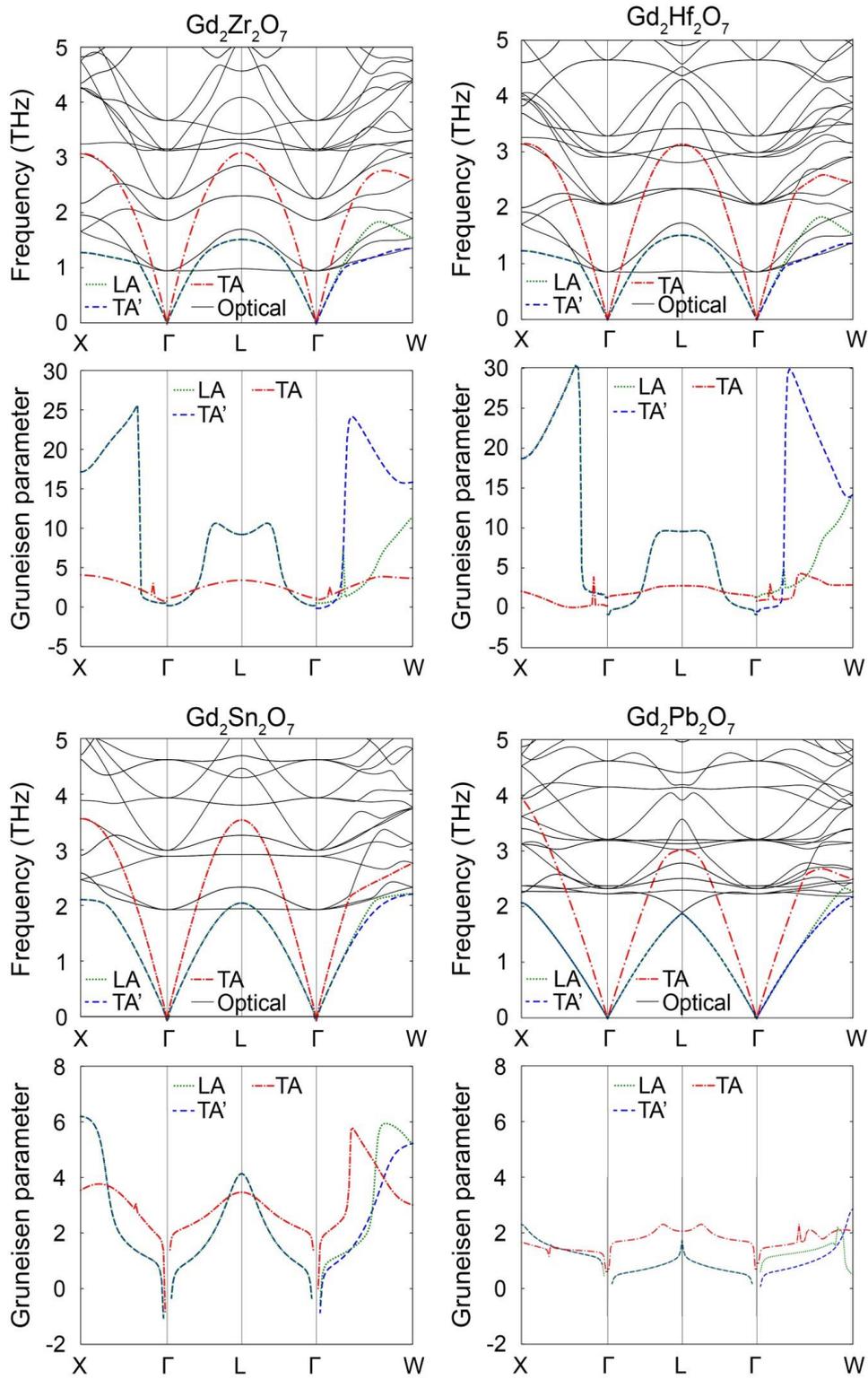

**Fig. 5** (Color online): Calculated phonon dispersions (zoom in for the acoustic branches, i.e., TA, TA',LA and optical branches, indicated by green dotted, blue dashed, red dashed-dotted and solid black lines, respectively) and corresponding acoustic Grüneisen parameters for $Gd_2T_2O_7$ (T = Zr, Hf, Sn, Pb) pyrochlores.

### 3.3. Prediction of thermal conductivities

Table IV shows the thermal conductivities predicted by the RTA model (cf. Eq. 3), the minimum thermal conductivities ($\kappa_{min}$) estimated from Clarke's and Cahill's models, and the experimental data. For the $Ln_2Zr_2O_7$ pyrochlores, we see that the predicted values of $\kappa_{min}$ from Clarke's and Cahill's models are almost the same, which is expected given the similar lattice constants, moduli (and thus sound velocities) and densities across the pyrochlores. This is apparently inconsistent with the trend revealed in the experimental data [15, 17, 19, 67], mostly showing an order of $\kappa_{La_2Zr_2O_7} > \kappa_{Nd_2Zr_2O_7} > \kappa_{Sm_2Zr_2O_7} > \kappa_{Gd_2Zr_2O_7}$ [68]. On the other hand, our predictions using the RTA model shows good agreement in both the values of $\kappa$ and trend with the experimental data for the $Ln_2Zr_2O_7$ pyrochlores. Additionally we can see from the data shown in Table IV that only LA branches in $Ln_2Zr_2O_7$ pyrochlores are effective to carry heat while the two transverse branches yield negligible contributions (<0.1 W m$^{-1}$ K$^{-1}$) to the thermal conductivity. The trivial contribution to $\kappa$ from the transverse branches directly derive from the large values of $\bar{\gamma}_{TA}$ and $\bar{\gamma}_{TA'}$ (cf. Table III). Meanwhile it is worth noting that, though much smaller than $\bar{\gamma}_{TA}$ and $\bar{\gamma}_{TA'}$, $\bar{\gamma}_{LA}$ exhibits a sizable magnitude (2-3), higher than the usual average Grüneisen parameters (often being 1-2) of most other systems [69].

Within the $Gd_2T_2O_7$ pyrochlore group, the LA branches also dominate the thermal conductivity for $Gd_2Zr_2O_7$ and $Gd_2Hf_2O_7$. Particularly for $Gd_2Hf_2O_7$, it has a low value of $\bar{\gamma}_{LA}$ (i.e., 1.99) despite very large values of $\bar{\gamma}_{TA}$ and $\bar{\gamma}_{TA'}$ (both beyond 8), and thus yield a relatively high $\kappa$ value. Unlike $Gd_2Zr_2O_7$ and $Gd_2Hf_2O_7$, $Gd_2Pb_2O_7$ and $Gd_2Sn_2O_7$ exhibit much lower $\bar{\gamma}_{TA}$ and $\bar{\gamma}_{TA'}$ values, even lower than $\bar{\gamma}_{LA}$. As a consequence, the TA/TA' branches of $Gd_2Pb_2O_7$ and $Gd_2Sn_2O_7$ become important in heat conduction though their contributions, i.e., $\kappa_{TA}$ and $\kappa_{TA'}$, remain smaller than the one by the LA branch, $\kappa_{LA}$ because of their smaller sound velocity and reduced Debye temperature. One thing to note is that $Gd_2Pb_2O_7$, with overall low average Grüneisen parameters, is predicted to yield the highest $\kappa$ (i.e., 3.65 W m$^{-1}$ K$^{-1}$) among all these pyrochlores considered according to the RTA model. This is in sharp contrast with the predictions from Clarke's and Cahill's model, both suggesting that $Gd_2Pb_2O_7$ have the lowest $\kappa$ because of its small phonon velocities and moduli. Another thing to note is that the $Gd_2Sn_2O_7$ exhibit largest $\bar{\gamma}_{LA}$ and thus yield a rather small $\kappa$ albeit somewhat small $\bar{\gamma}_{TA}$ and $\bar{\gamma}_{TA'}$.

In Figure 6, we compare the predictions of $\kappa$ from the RTA model with the available experimental data over the temperature range of 600K-1600K for the RE pyrochlores (abate $Gd_2Hf_2O_7$ and $Gd_2Pb_2O_7$ due to absence of experimental data). In general we can see that our predictions agree well with the experimental data. However there are some discrepancies between the predicted and experimentally measured $\kappa$ values at low (i.e., 600K $<$ $T$ $<$ 800K) and high (i.e., $T$ $>$ 1400K) temperatures for the $Ln_2Zr_2O_7$ pyrochlore group. These discrepancies are because the Umklapp scattering is assumed to dominate the thermal resistance. Particularly in the low temperature regime, the predicted $\kappa$ values are generally higher than the experimental measurements, i.e., the RTA model overestimates the thermal conductivity. This overestimation may come from the fact that the samples in experiments are polycrystalline with considerable amount of lattice defects that induce extra phonon scattering [32] to lower the thermal conductivity. In the high temperature regime, the experimental $\kappa$ values exhibit slower downward trend or even slightly increase, which can be due to high-temperature radiation that provides positive contribution to thermal conductivity [32] and thereby offsets the downward trend attributed to Umklapp scattering processes. One exception in Fig. 6 is the case of $Gd_2Sn_2O_7$ where the RTA model prediction is shown to capture the correct temperature dependence of $\kappa$ but somewhat underestimate the magnitude. However the statistical significance of the above comparison for $Gd_2Sn_2O_7$ is questionable given the insufficient experimental data (cf. Fig. 6f).

To compare the thermal conductivities between different RE pyrochlores, the predicted $\kappa$ versus temperature curves (from the RTA model) are plotted together for the $Ln_2Zr_2O_7$ and $Gd_2T_2O_7$ pyrochlore groups in Figs 7a and 7b respectively. Furthermore, we define a relative ratio $\rho_\kappa$ of thermal conductivity, taking $Gd_2Zr_2O_7$ as the reference:

$$r_k = \frac{k_{Ln_2T_2O_7} - k_{Gd_2Zr_2O_7}}{k_{Gd_2Zr_2O_7}},$$

(10)

where $k_{Ln_2T_2O_7}$ and $k_{Gd_2Zr_2O_7}$ are the RTA model predicted thermal conductivities for $Ln_2T_2O_7$ (Ln = La, Nd, Sm, Gd and T = Zr, Hf, Sn, Pb) and $Gd_2Zr_2O_7$ respectively. Fig. 7c and 7d show the plots of $\rho_\kappa$ versus the $Ln^{3+}$ and $T^{4+}$ cation radius respectively with the error bar indicating the standard deviation of $\rho_\kappa$ with respect to the temperature The values of $\rho_\kappa$ for experimental data were also calculated, in the same way as Eq. 10 but with the predicted values replaced by experimental values. Here only those experimental studies that contain measurements of $Gd_2Zr_2O_7$ and other pyrochlores simultaneously were considered, to avoid errors due to experimental methods and samples. The usage of the relative ratio $\rho_\kappa$ helps moderate the radiated and spurious effects during to scattered data in experiments to enable a more accurate and quantitative comparison between the experimental measurements with our predictions. For the $Ln_2Zr_2O_7$ pyrochlore group where experimental data are available, the predicted and experimental $\rho_\kappa$ values show a close match (cf. Fig. 7c). This good agreement further evidences that the RTA model captures the essential physics underlying thermal transport in RE pyrochlores. In addition, there is an apparent increasing trend in $\rho_\kappa$ as the $Ln^{3+}$ cation radius increases. For the $Gd_2T_2O_7$ pyrochlore group, the experimental $\rho_\kappa$ values are not available. Our predictions suggest that, when measured under same experimental conditions, $Gd_2Sn_2O_7$ would yield a similar $\kappa$ value as $Gd_2Zr_2O_7$, but much smaller than the ones of $Gd_2Hf_2O_7$ and $Gd_2Pb_2O_7$, thus being a potential TBC material like $Gd_2Zr_2O_7$.

**Table III**: Average longitudinal ($v_{LA}$) and transverse ($v_{TA/TA'}$) phonon velocities, average Grüneisen parameters $\bar{\gamma}_i$ ($i$ = TA, TA' or LA), reduced Debye temperatures ($\widetilde{\Theta}_{LA}$ and $\widetilde{\Theta}_{TA/TA'}$) calculated from phonon dispersions (cf. Figs 4-5). The phonon velocity is taken as the slope of each acoustic branch close to the $\Gamma$ point while the reduced Debye temperature is computed using Eqs 5-6.

|  | $\bar{\gamma}_{LA}$ | $\bar{\gamma}_{TA}$ | $\bar{\gamma}_{TA'}$ | $v_{LA}$ | $v_{TA/TA'}$ | $\widetilde{\Theta}_{LA}$ | $\widetilde{\Theta}_{TA/TA'}$ |
|---|---|---|---|---|---|---|---|
| $La_2Zr_2O_7$ | 2.48 | 4.87 | 4.35 | 5878 | 3125 | 256 | 136 |
| $Nd_2Zr_2O_7$ | 2.75 | 6.23 | 10.10 | 5906 | 3126 | 260 | 138 |
| $Sm_2Zr_2O_7$ | 2.81 | 7.40 | 11.98 | 5783 | 3086 | 256 | 137 |
| $Gd_2Zr_2O_7$ | 2.75 | 7.19 | 11.57 | 5634 | 3059 | 252 | 137 |
| $Gd_2Hf_2O_7$ | 1.99 | 8.07 | 12.53 | 5105 | 2787 | 229 | 125 |
| $Gd_2Pb_2O_7$ | 1.73 | 1.18 | 1.06 | 4463 | 2450 | 195 | 107 |
| Gd$_2$Sn$_2$O$_7$ | 3.19 | 2.68 | 2.46 | 5388 | 3178 | 242 | 142 |

**Table IV**: Calculated minimum thermal conductivity ($\kappa_{min.}$, W m$^{-1}$ K$^{-1}$) from Clarke's and Cahill's model; Calculated thermal conductivity ($\kappa$, W m$^{-1}$ K$^{-1}$) from the RTA model (cf. Eqs 3-4) at 1073 K; Experimental thermal conductivity at 1073 K ($\kappa_{Exp.}$, W m$^{-1}$ K$^{-1}$).

| | $\kappa_{min.}$ | | $\kappa$ from the RTA model at 1073 K | | | | $\kappa_{Exp.}$ at 1073 K | | | |
|---|---|---|---|---|---|---|---|---|---|---|
| | Clarke's | Cahill's | LA | TA | TA' | Total | Liu[17] | Lehmann[19] | Suresh[15] | Pan[67] |
| La$_2$Zr$_2$O$_7$ | 1.21 | 1.14 | 1.58 | 0.06 | 0.07 | 1.72 | | 1.55 | 1.43 | |
| Nd$_2$Zr$_2$O$_7$ | 1.23 | 1.16 | 1.36 | 0.04 | 0.01 | 1.41 | 1.48 | 1.24 | | |
| Sm$_2$Zr$_2$O$_7$ | 1.23 | 1.16 | 1.26 | 0.03 | 0.01 | 1.30 | 1.44 | | | 1.20 |
| Gd$_2$Zr$_2$O$_7$ | 1.22 | 1.15 | 1.26 | 0.03 | 0.01 | 1.30 | 1.33 | | 1.03 | 1.18 |
| Gd$_2$Hf$_2$O$_7$ | 1.07 | 1.06 | 2.31 | 0.02 | 0.01 | 2.34 | | | | |
| Gd$_2$Pb$_2$O$_7$ | 0.92 | 0.88 | 2.06 | 0.71 | 0.88 | 3.65 | | | | |
| Gd$_2$Sn$_2$O$_7$ | 1.17 | 1.16 | 0.90 | 0.25 | 0.30 | 1.45 | | | | |

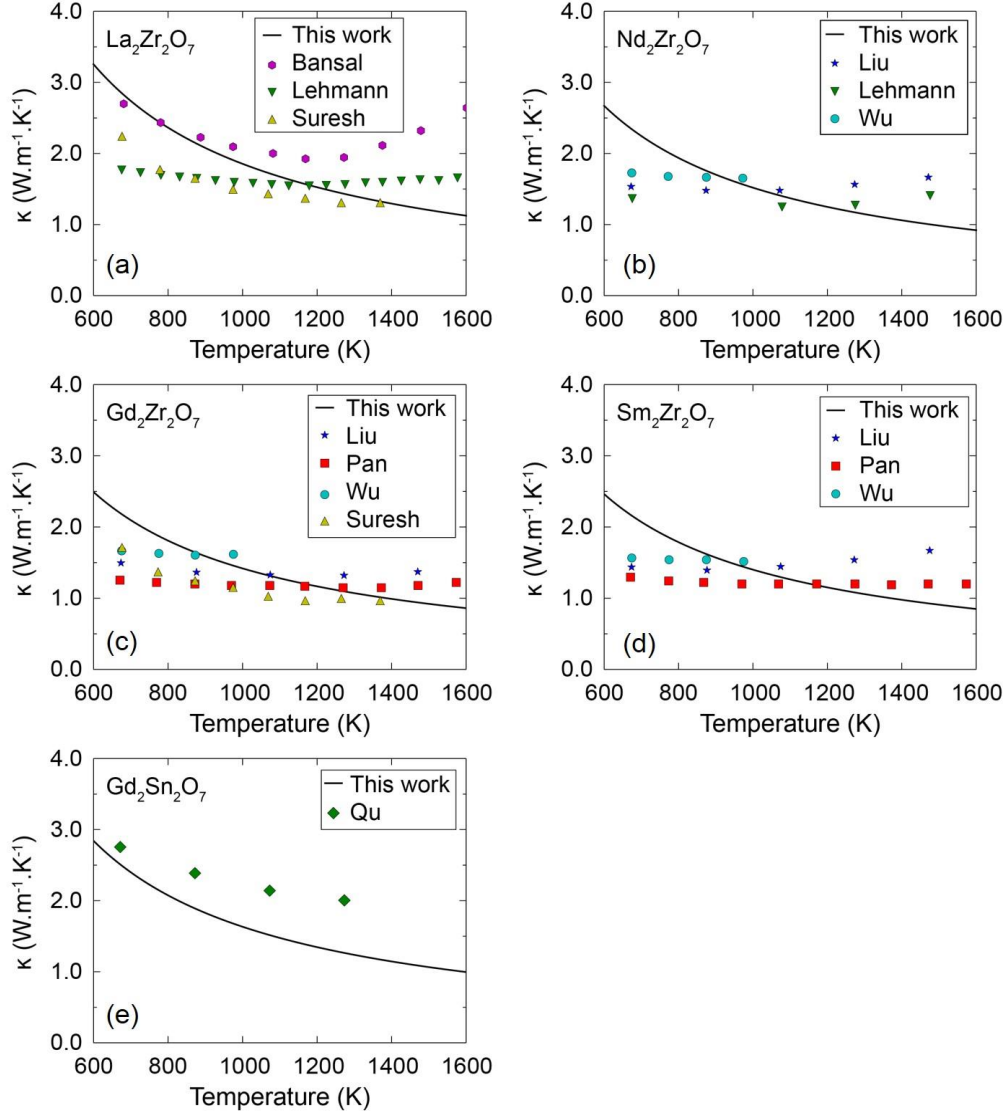

**Fig. 6** (Color online): Lattice thermal conductivities ($\kappa$) of $Ln_2Zr_2O_7$ (Ln = La, Nd, Sm, Gd) and $Gd_2Sn_2O_7$. The solid red lines represent the predicted values of $\kappa$ using the RTA model. The various filled symbols represent the experimental measurements from Bansal *et al.* [14] (hexagon), Lehmann *et al.* [19] (down triangle), Suresh *et al.* [15] (up triangle), Liu *et al.* [17] (star), Pan *et al.* [67] (square), Wu *et al.* [16] (circle) and Qu *et al.* [70] (diamond), respectively. The experimental data for $Gd_2Hf_2O_7$ and $Gd_2Pb_2O_7$ are missing.

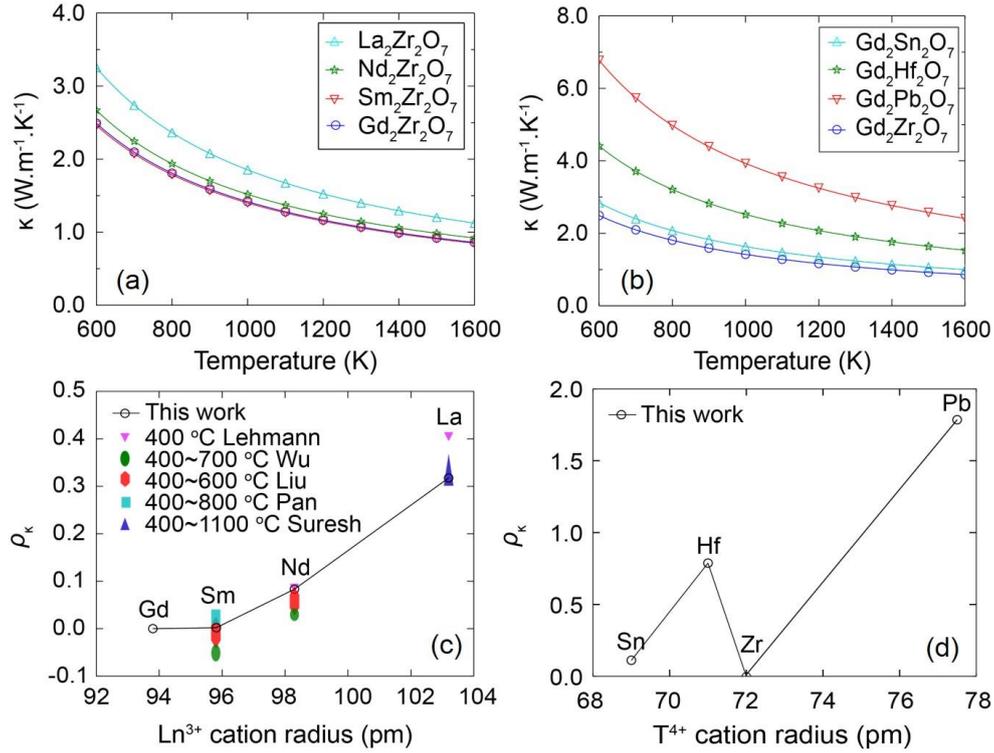

**Fig. 7** (Color online): Predicted thermal conductivities ($\kappa$) for (a) $Ln_2Zr_2O_7$ and (b) $Gd_2T_2O_7$ pyrochlore groups respectively. The relative ratio of thermal conductivity ($\rho_\kappa$) taking $Gd_2Zr_2O_7$ as the reference (cf. Eq. 10), are plotted versus the (c) $Ln^{3+}$ and (d) $T^{4+}$ cation radius respectively.

### 3.4. Structural origin of low-lying optical branches

Clearly evidenced from our results above, the low-lying optical phonon branches play a critical role in determining the thermal conductivities of RE pyrochlores. They strongly scatter the acoustic branches, leading to extremely large Grüneisen parameters (cf. Figs 4-5 and Tables III) and subsequently low $\kappa$ values. As we already know from Section 3.2.1, the low-lying optical branches are attributed to the vibrations of the Ln atom. A close examination of the vibrational patterns responsible for the low-lying optical branches reveals that Ln exhibits sizable vibrational magnitude while the O' (8b) atom shows near zero vibration, and T and O (48f) atoms show very small amplitude of vibration, mostly lesser than 15% of that of the Ln atom.

Particularly for the Ln atom, there exist two possible modes of vibration associated with the low-lying optical branches at $X$, $\Gamma$ and $L$ points, one being that Ln moves in-plane (i.e., on the perpendicular to the O'- Ln bond, cf. Fig. 8) towards the adjacent O (48f) atom while the other being that Ln is displaced toward to one of the nearest T atoms, as illustrated in Fig. 8.

The large vibrational magnitude of Ln is a direct consequence of loose bonds between Ln and its adjacent atoms. In this regard, we examined the charge distribution around the Ln atom to further elucidate its bonding environment. Since the bonds within the RE pyrochlores mostly exhibit ionic features, the Voronoi deformation density (VDD) method [71] was employed to analyze atomic static charges. Compared to other charge-analyzing methods, the VDD method has the advantage of giving chemically meaningful charges with basis-set independency [71]. The calculated static charges using VDD for Ln, T, O (48f) and O' (8b) atoms are presented in Fig. 9a and Fig. 9b for $Ln_2Zr_2O_7$ (Ln = La, Nd, Sm, Gd) and $Gd_2T_2O_7$ (T = Zr, Hf, Sn, Pb) pyrochlores respectively.

For the $Ln_2Zr_2O_7$ pyrochlore group, the charge at the Ln site increases as the cation radius increases while in accordance the charges at O (48f) and O' (8b) become more negative. This indicates that Ln loses electrons more easily (to its neighboring O (48f) and O' (8b) atoms) as the cation radius increases. However we can see from Fig. 9a that the charge change at O (48f) is rather negligible, suggesting that the charge transfer mainly occur between Ln and O' (8b), which is not surprising with O' (8b) sites being the nearest neighbors to Ln. Apparently the net increase in charge transfer from Ln to surrounding atoms may contribute to enhancing the local ionic interactions. This provides possible explanation for the elevation of optical branches and lower $\bar{g}_{TA}/\bar{g}_{TA'}$ of $La_2Zr_2O_7$ in comparison to other $Ln_2Zr_2O_7$ pyrochlores. Meanwhile the charge at Zr remains largely invariant irrespective of the species of Ln. This together with the minimal change in the charges of O (48f) sites indicates that the bonding environment around Zr remains mostly unchanged, and thus explains the invariance of the branches corresponding to $Zr^{4+}$ shown in Fig. 2.

Meanwhile for the $Gd_2T_2O_7$ pyrochlore group (Fig. 9b), the Zr and Hf sites exhibit charge values ~3 while the Sn and Pb sites exhibit charge values close to 2, and accordingly the O (48f) sites neighboring $Gd_2Sn_2O_7$ and $Gd_2Pb_2O_7$ show less negative charges compared to the ones in $Gd_2Zr_2O_7$ and $Gd_2Hf_2O_7$. These large discrepancies in charge values directly evidence the less ionicity but more covalency in the characteristics of T-O bonds in $Gd_2Sn_2O_7$ and $Gd_2Pb_2O_7$ than $Gd_2Zr_2O_7$ and $Gd_2Hf_2O_7$, as previously mentioned in Section 3.2.1. The above difference is also well reflected in Fig. 10 where we see the charge center of O (48f) anion stay close to its neighboring T cations in the cases of T= Zr and Hf while drift towards the Gd cation in the cases of T= Sn and Pb. Although from the charges at Ln and O' (8b) are barely affected by the presence of different T atoms (cf. Fig. 9b), the proximity of O (48f) charge centers to Gd in $Gd_2Sn_2O_7$ and $Gd_2Pb_2O_7$ will strength the Gd-O (48f) bonds. The stronger bonding explains the elevation of the low-lying optical branches (corresponding to Gd vibrations) and significantly smaller $\gamma_{TA}/\gamma_{TA'}$ for $Gd_2Sn_2O_7$ and $Gd_2Pb_2O_7$ (cf. Fig. 4).

The above structural origin underlying low-lying optical branches provides intuitive insights towards understanding lattice thermal conductance in RE pyrochlores. For instance, $Gd_2Pb_2O_7$ has the lowest overall bond strength (moduli) among those RE pyrochlore systems but meanwhile has strong Gd-O (48f) bonds which induce elevated low-lying optical branches that have less interference with the transport of acoustic phonons. As such, $Gd_2Pb_2O_7$ exhibits a relatively high thermal conductivity. Nonetheless, it is worth pointing out that the elevation of low-lying optical branches does not always favor the transport of acoustic phonons. As seen in the case of $Gd_2Sn_2O_7$, though inducing less scattering in the TA/TA' branches, the elevated optical branches cause considerable scattering in the LA branch which is of higher frequency than TA branches, leading to a rather low thermal conductivity (cf. Fig. 5 and Table IV). Thus further studies are necessary to clarify the competition between TA/TA' scattering and LA scatting, to fully understand the role of low-lying optical branches in thermal conductance in RE pyrochlores.

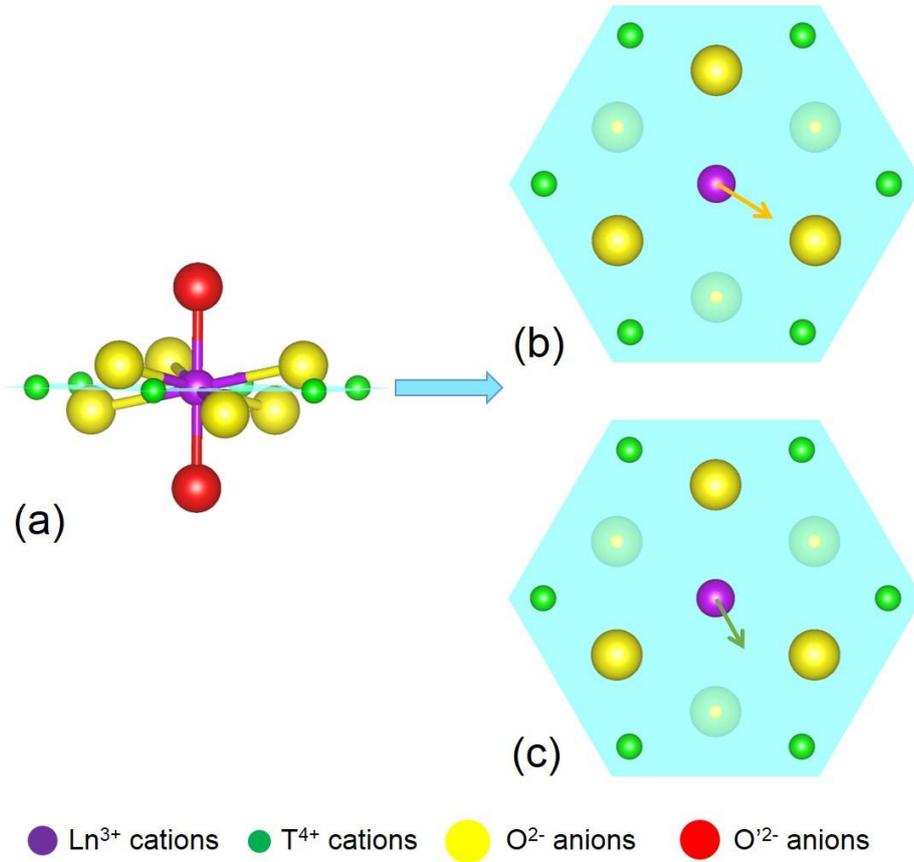

**Fig. 8** (Color online) (a) Schematic showing the bonding environment around the Ln (16$d$) atom, showing nearest six O (48$f$), two O' (8$b$) and six Zr (16$c$). The cyan plane indicates the plane that cuts through Ln and is perpendicular to the O'- Ln bond. The projection views of the two possible vibrational motions (associated with low-lying frequencies) of the Ln atom, i..e, (b) Ln moving in-plane towards the adjacent O (48f) atom, and (c) Ln moving toward to one of the nearest T atoms, are illustrated with vibrational directions indicated by the yellow and green arrows respectively.

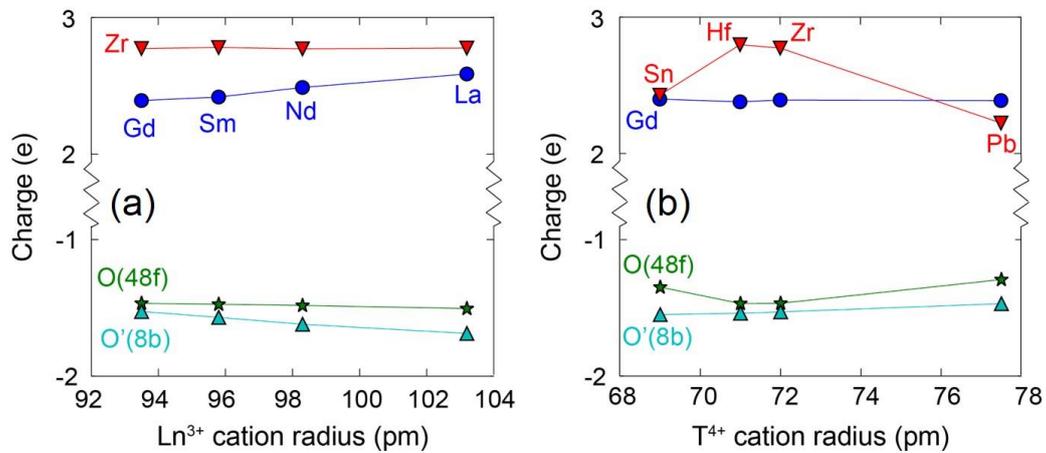

**Fig. 9** (Color online): Calculated static charges at different atom sites (with the VDD method) for (a) $Ln_2Zr_2O_7$ and (b) $Gd_2T_2O_7$ pyrochlores, respectively.

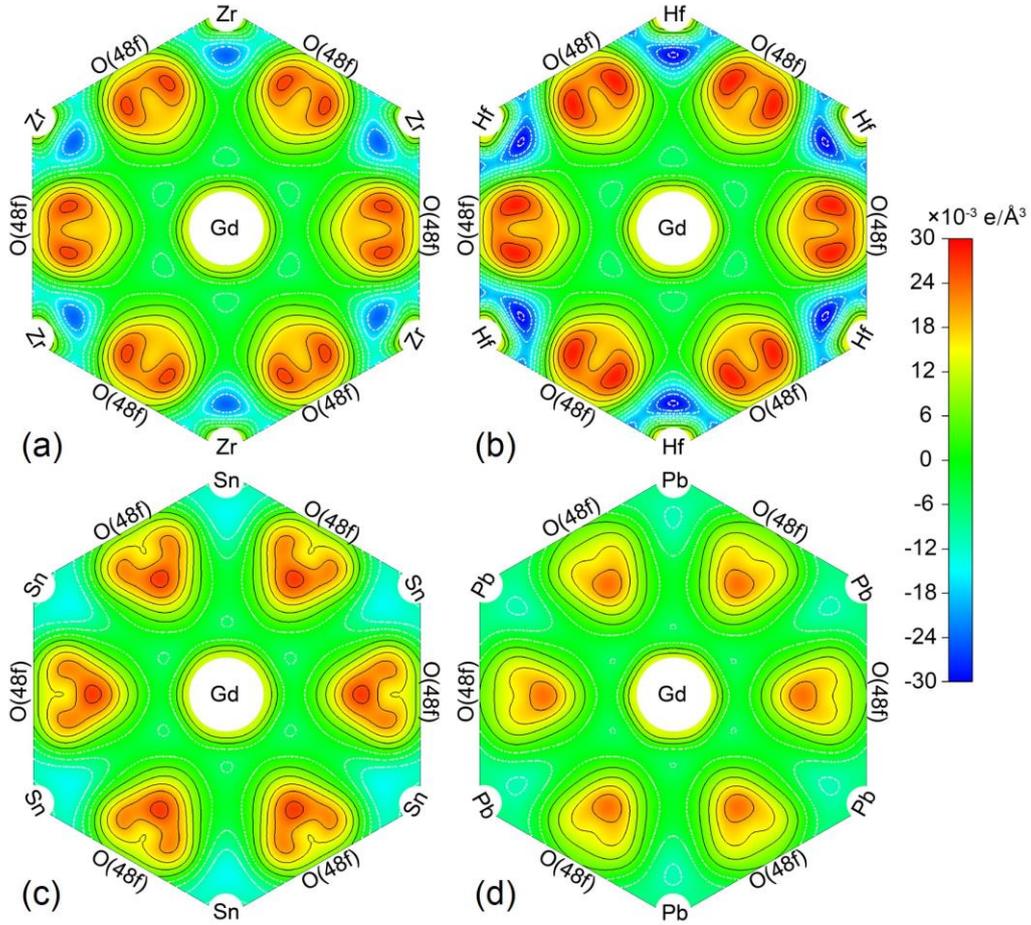

**Fig. 10** (Color online) Calculated deformation charge density contours on the planes cutting through Gd and perpendicular to the O'- Gd bond for (a) $Gd_2Zr_2O_7$, (b) $Gd_2Hf_2O_7$, (c) $Gd_2Sn_2O_7$ and (d) $Gd_2Pb_2O_7$ pyrochlores, respectively. The contours are taken at intervals of 0.005 $e/Å^3$. The solid black contours and red color represent positive values (gaining electrons), while the white dash contours and blue color correspond to negative values (losing electrons).

## 4. Summary

In summary, the phonon properties and thermal conductivities of two sets of RE pyrochlores, $Ln_2Zr_2O_7$ (Ln = La, Nd, Sm, Gd) and $Gd_2T_2O_7$ (T = Zr, Hf, Sn, Pb) have been studied through density functional theory calculations and quasi harmonic approximation, combining the relaxation-time approximation (RTA) together with the Debye model. Compared to the often-used minimum thermal conductivity approach using Clarke's and Cahill's models that hardly differential different RE pyrochlores, the RTA with the Debye model yields much more accurate predictions of thermal conductivities of RE pyrochlores, validated by available experimental measurements. In particular, for the $Ln_2Zr_2O_7$ pyrochlores, the thermal conductivity is predicted to increase as the $Ln^{3+}$ cation radius increases, in close agreement with experiments.

Our results indicate that the low thermal conductivities of RE pyrochlores mainly derive from the interference between the low-lying optical branches and acoustic branches. For the $Ln_2Zr_2O_7$ pyrochlore group and $Gd_2Hf_2O_7$, the low-lying optical phonon branches cause significant scattering of the transverse acoustic branches, leading to very large Grüneisen parameters and rendering the transverse branches largely irrelevant in lattice thermal conductance. In comparison, $Gd_2Pb_2O_7$ and $Gd_2Sn_2O_7$ exhibit much elevated low-lying optical phonon branches and therefore much less scattering in the transverse branches. Consequently the transverse branches produce decent contributions to the thermal conductivity. The above interplay between the low-lying optical branches and transverse branches in different RE pyrochlores, i.e., $Ln_2T_2O_7$ (Ln = La, Nd, Sm, Gd, T = Zr, Hf, Sn, Pb), are shown to originate from the bonding between the Ln and its adjacent O atoms where the bonding strength is affected by the charge transfer between Ln and O' (8b) atoms as well as the shifting of the O (48f) charge centers. Meanwhile, our study also indicated that the low-lying optical phonon branches, when elevated, may induce substantial scattering in the longitude acoustic branch of the RE pyrochlores, suggesting that further studies are necessary to clarify the playoff between scattering processes in transverse and longitude acoustic branches in order to fully elucidate the role of low-lying optical branches in determining the thermal conductivity in RE pyrochlores.

## Acknowledgement

We greatly thank the financial support from McGill Engineering Doctoral Award and National Sciences and Engineering Research Council (NSERC) Engage grant (grant # EGP #452480-13) and Discovery grant (grant # RGPIN 418469-2012). We also acknowledge Supercomputer Consortium Laval UQAM McGill and Eastern Quebec for providing computing power.